\documentclass[prc,aps,floatfix,nofootinbib,twocolumn,superscriptaddress]{revtex4} 

\usepackage{graphicx}
\usepackage{color}
\usepackage{enumerate}
\usepackage{amssymb}
\usepackage{amsmath}
\usepackage{amsfonts}
\usepackage{bbm}
\usepackage{dsfont}
\def\be{\begin{equation}} 
\def\ee{\end{equation}}

\newcommand{\mat}[1]{\mbox{\boldmath $#1$}}

\begin{document}

\author{G.F. Bertsch}
\affiliation{ 
Department of Physics and Institute for Nuclear Theory, Box 351560, 
University of Washington, Seattle, Washington 98915, USA}

\author{K. Hagino}
\affiliation{ 
Department of Physics, Kyoto University, Kyoto 606-8502,  Japan} 

\title{Barrier penetration in a discrete-basis formalism }

\begin{abstract}
The dynamics of a many-particle system are often modeled by mapping the
Hamiltonian onto a Schr\"odinger equation.
An alternative approach is to solve the Hamiltonian equations
directly in a model space of many-body configurations. In a previous paper
the numerical convergence of the two approaches was compared with a simplified
treatment of the Hamiltonian representation.  Here we extend the comparison
to the nonorthogonal model spaces that  would be obtained by the 
generator-coordinate method.  With a suitable choice of the
collective-variable grid, a configuration-interaction Hamiltonian can
reproduce the Schr\"odinger dynamics very well.
However, the method as implemented here requires that the
barrier height is not much larger than the zero-point energy in
the collective coordinates of the configurations.
\end{abstract}

\maketitle

\section{Introduction}

Ever since the pioneering work of Hill and Wheeler \cite{HW},   
low-energy fission has been parameterized by their  barrier
penetration formula,  based on a one-dimensional Schr\"odinger equation.
See Refs. \cite{si16,ko23,be23,kawano2023} for recent extensions of that model.
A kinetic energy
operator and  potential energy function for the Schr\"odinger equation
can be derived in the Generator Coordinate Method of many-body
theory, but beyond the extension to two dimensions\cite{2D} the generalization to
other degrees
of freedom presents formidable obstacles \cite{GY}.  In contrast,
Hamiltonians constructed from the configuration-interaction (CI) approach 
\cite{al2020,BY,BH2022,BH2023} can in principle deal with any mechanisms 
present in nuclear dynamics.  

A CI basis is usually constructed from nucleonic Hamiltonians 
by solving the Hartree-Fock or Hartree-Fock-Bogoliubov equations
in the presence of shape constraints.  Those constraints map out a path for the barrier
crossing.    To proceed further without leaving the CI basis 
one needs to understand how to deal with the non-orthogonality of the 
constrained configurations. Also, as a practical question, how closely do
the configurations need to be spaced along a fission path to reproduce the
Schr\"odinger 
dynamics? In this work we apply reaction theory as
formulated in a discrete basis of states 
to investigate how well that framework can reproduce the Schr\"odinger. 

The focus of this study is the transmission probability $T$ for traversing
an isolated  one-dimensional barrier, following up on the work of 
Ref. \cite{ha23}.  We  assume that the 
barrier potential vanishes at large distances,  so the wave function
satisfies ordinary plane-wave boundary conditions.

\section{Model space and Hamiltonian}

\subsection{Basics}
The construction of the model space and the Hamiltonian within it closely follows
the treatment in Ref. \cite{BY}.  The states in the space
are obtained by self-consistent mean-field theory augmented by a
$q$-dependent constraining field. A finite basis is generated 
on a mesh of points $\{q_i\}$, making a path along the collective coordinate.  
With those wave functions
one computes the overlaps
\be
\mat{N}_{ij} = \langle \psi_{i} | \psi_{j} \rangle.
\ee
Here and hereafter, we use boldface symbols for matrices.
The Hamiltonian matrix elements are similarly computed with a Hamiltonian
$H$ that contains a nucleon-nucleon interaction,
\be
\mat{H}_{ij} = \langle \psi_{i} |  H | \psi_{j} \rangle.
\ee 
Insight into the workings of this approach can be obtained by taking the
center-of-mass coordinate  as a paradigm of a collective variable\cite{BY}. One finds
that $\mat{N}_{ij}$ can be parameterized quite well as a Gaussian,
\be
\mat{N}_{ij} \approx n_{|i-j|} = \exp\left( - (q_i - q_j)^2/ 4 s^2\right)
\ee 
where $s$ is a physical parameter associated with the size of the collective
wave packets.  The states giving rise to the above overlaps
have a separable form
\be
\label{eq:psiq}
\psi_{i}(q,\vec{\xi}) = \psi_{\rm int}(\vec{\xi}) \exp\left(-(q-q_i)^2/2 s^2\right)
\ee
where $\psi_{\rm int}$ depends only on intrinsic coordinates $\vec{\xi}$.
One assumes that the Hamiltonian can be separated into an intrinsic
part and a collective kinetic part given by  
\be
\hat H^0 =  - \frac{\hbar^2}{2M_q}\frac{\partial^2}{\partial q^2}.
\label{eq:H0}
\ee
Here $M_q$ is an inertial parameter associated with the collective
coordinate.  The  matrix elements of $\hat H^0$ are parameterized as
\be
\mat{H}^0_{ij} \approx  h_{|i-j|} = n_{|i-j|} E_q \left(1 - (q_i - q_j)^2/2 s^2\right)
\label{eq:Eq}
\ee
where $E_q = \hbar^2/(4 M_q s^2)$ is the zero-point energy of the configuration.

In using a discrete-basis representation in reaction theory, it is helpful to understand how
to represent noninteracting plane waves.   For this purpose, we
choose a grid of uniformly spaced points separated by
$\Delta q = q_{i+1} - q_i $.  The eigenstates of $H^0$ 
are given by
\be
\Psi_k (q,\vec{\xi})= \sum_n    \psi_{n}(q,\vec{\xi}) \,r^n, 
\label{eq:psik}
\ee
where $r = e^{i k \Delta q}$ and $k$ is a momentum index.
Note that the space states only support momenta in the
range $ -\pi <   k \Delta q  < \pi$.  

The kinetic energy $E_{DB}$ of the plane-wave state in the discrete basis
is given by \cite{BY}
\be
E_{DB}(k)  = \frac{ h_0 + 2 \sum_{j>0} h_j \cos{j k\Delta q}}
{ 1 + 2 \sum_{j>0} n_j }.
\label{eq:Egcm}  
\ee
In practice $\mat{H}$ will be treated as a band-diagonal matrix with
matrix elements $\mat{H}_{ij}$ set to zero for $|i-j| > N_{od}$.
In the simplest version of the theory, $N_{od}=1$ and the matrix
is tridiagonal with interactions only between nearest neighbors.
The quality of the energy fit to
the Sch\"odinger energy $E_{s} = k^2\hbar^2/2 M_q$ depends on $N_{od}$
and on the dimensionless ratio $\Delta q / s$.  The computed  $E_{DB}(k)$
does not go exactly to zero at $k=0$, since the sum of the Gaussians
in Eq. (\ref{eq:psik}) still has some variation as a function of $q$.
To keep the energy comparisons consistent, we
compare the excitation energy
\be
E = E_{\rm DB}(k)-E_{\rm DB}(0)
\ee 
with the Schr\'odinger energy $E_s$.

Fig. 1 shows the comparison with some examples.  
In Ref. \cite{BY} the choice $\Delta q = 5^{1/2}s$ 
was advocated for discrete-basis Hamiltonians of tridiagonal form.
The derived eigenenergies for the 
tridiagonal and next-to-nearest-neighbor approximations in the
range $ 0 < k < \pi/ s $ are shown
in Fig. 1(a).
One can see
small differences between the two,
but overall it appears that the tridiagonal approximation is acceptable
up to energies $ \sim 2 E_q$.
\begin{figure}[htb] 
\begin{center} 
\includegraphics[trim= 0   0  0  0,clip=true,width=0.8\columnwidth]{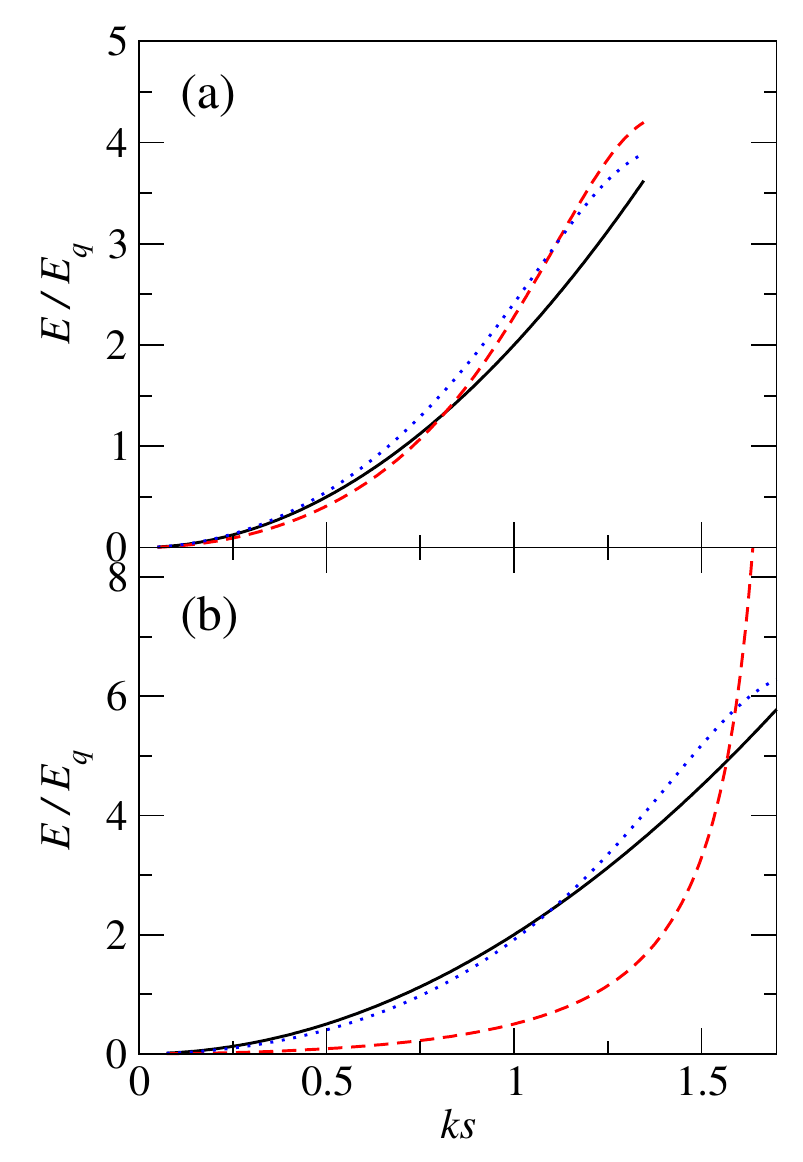} 
\caption{Energies  of plane-wave states in the discrete-basis formalism
for mesh parameters $\Delta q = 5^{1/2}$ (upper panel) and $2/3 \times 5^{1/2}$
(lower panel).  Black :  Schr\"odinger Hamiltonian $\hat H^0=k^2\hbar^2/2M_q$; red dashed:
discrete-basis Hamiltonian with  $N_{od} = 1$; blue dotted: $N_{od} = 2$.
Energies are in units of $E_q$, and momenta in units of $s$.
\label{fig:Ek}} 
\end{center} 
\end{figure} 
Fig. 1(b) shows the spectra for a somewhat smaller mesh spacing, 
 $\Delta q = \frac{2}{3}5^{1/2}\, s$.  Here the tridiagonal 
treatment fails.  On the other hand, inclusion of  next-to-nearest
neighbors ($N_{od} = 2$)   restores a good approximation to the
energy curve and increases the range of $k$.

\subsection{The barrier}

We consider the transmission coefficient for a plane
wave incident on a barrier of the form
\be
\label{eq:V}
V(q) = V_0 \exp\left( - q^2/2 \sigma^2\right).
\ee
This simulates a quadratic barrier around $q=0$ but vanishes
at large distances.  
The matrix elements for   wave functions in Eq. 
(\ref{eq:psiq}) are
\begin{eqnarray}
&&\langle \psi_{i} | \hat V(q) | \psi_{j}\rangle
= \mat{V}_{ij}  \nonumber \\
&=&V_0 \sqrt{\frac{2 \sigma^2}{s^2 + 2 \sigma^2}} 
\exp\left(-(q_1+q_2)^2/4(s^2 + 2 \sigma^2)\right)
\mat{N}_{ij}.\nonumber \\
\label{eq:Vij}
\end{eqnarray}

Besides $\Delta q$ and $N_{od}$, a third numerical parameter in the
discrete-basis formulation  is the dimension $N_{DB}$ 
of the $H$ and $N$ matrices.  We will see that the space 
needs to extend beyond the range of the barrier by only a few 
states to produce fairly accurate transmission probabilities.

\subsection{The transmission probability}

The energies $E$ and eigenstates 
$\vec{\psi}_E = (f_1,f_2,...,f_{N_{od}})$ 
of the matrix  
Hamiltonian $\mat{H}_{ij}  = \mat{H}^0_{ij}  +  \mat{V}_{ij} $
satisfy the equation 
\be
 \mat{H}'   \vec{\psi}_E \equiv (\mat{H}-\mat{N}\,E)\vec{\psi}_E = 0
\label{eq:H'}
\ee
for rows $m$ that contain all of the possible elements
of $\vec{H}'$ in its band-diagonal construction.
The
rows beyond  $ m = N_{\rm grid} -N_{od}$ lack one or more matrix elements and
must be modified to insure that the transmitted
wave satisfies an outgoing-wave boundary condition.  The same applies to
the topmost rows with  $m < N_{od}$.  The boundary condition here
requires the wave function to be composed
of a linear combination of incoming and reflected plane waves.  This
is achieved by modifying the diagonal $\mat{H}'_{mm}$  and replacing
Eq. (\ref{eq:H'}) by the inhomogeneous
equation 
\be
(\mat{H}' + \mat{\Delta H}')\vec{\psi}_E = \vec{v}.
\label{eq:H''v}
\ee
The general expressions  for $\mat{\Delta H}'$ and the vector $\vec{v}$ are derived in
the Appendix.  Following the numerical solution of Eq. (\ref{eq:H''v})
the transmission probability\footnote{This  method is 
an alternative to standard $S$-matrix theory\cite{mi93,da95,mi10}.
The present approach avoids the necessity of calculating the real and
imaginary parts of the coupling between states  in $\mat{H}$  
and the scattering channels.}
is evaluated as
\be
T_{DB} = 1- \frac{|f_1 e^{ik \Delta q}-f_2|^2}
{|f_1 e^{-i k \Delta q}-f_2|^2}.
\label{eq:T_CI}
\ee
In previous publications (e.g., Ref. \cite{BH2023}) we examined the theory at the level
of the $N_{od} = 1$ approximation.  
In this work  we also consider the $N_{od} = 2$ approximation.
The Appendix also includes the detailed formulas for this case.

The discrete-basis formalism defined in this way satisfies an
important check on the theory.  Eq. (\ref{eq:H''v}) can be solved analytically if
$\mat{V}$ vanishes, yielding the plane-wave solution 
$\vec{\psi}_k$ of  Eq. (\ref{eq:psik}) 
with  $k$ satisfying $E = E_{DB}(k)$.  Thus $T_{DB}=1$ 
trivially when there is no barrier.
     
One should be cautious in using the discrete-basis formalism at higher
energies even though they may still be in the allowed range of $E_{DB}(k)$.
As an extreme example, the energy is
a maximum at $k\,\Delta q = \pi$  but the corresponding wave function
is a pure standing wave with amplitudes $\vec{\psi}_i$ alternating in sign.
Its transmission probability is zero. This is easily seen from Eq. (\ref{eq:T_CI}).  The transmission
probability reaches a maximum somewhere inside the allowed range of
$E_{DB}$ and decreases at higher energies (see Fig. 4 below).  Obviously the 
treatment is then unphysical.

\section{Numerical examples}

We now compare calculated transmission probabilities $T_{DB}$ 
with those obtained by integrating
the Schr\"odinger equation\footnote{The resulting $T_s$ is
quite close to the Hill-Wheeler transmission probability 
$T_{HW} = [ 1 + \exp(- 2 \pi (E-V_0)/\hbar \omega)]^{-1}$ if 
the curvature parameter $\omega = (V_0 / \sigma^2 M_q)^{1/2}$ is the same.
A comparison is provided  in the Supplementary Material.}
\be 
\left(\hat H^0_q + V(q)\right)\phi = E \phi.
\label{eq:schrod}
\ee

\begin{figure}[htb] 
\begin{center} 
\includegraphics[trim= 0   0  0 0,clip=true,width=\columnwidth]{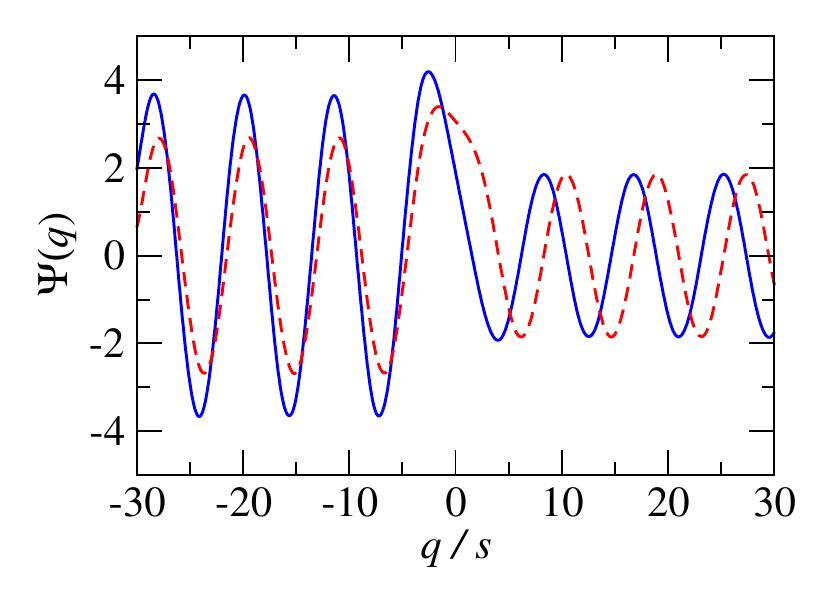} 
\caption{
Scattering wave function partially transmitted across a barrier.  Physical
parameters are 
$(V_0,\sigma,E) = (1.0,2.0,1.0)$  
in length and
energy units $s$ and $E_q$, respectively.  The numerical parameters are
$(N_{\rm grid},N_{od},\Delta q) = (42,2,2\times 5^{1/2}/3)$.  The real and imaginary parts
of the wave function are shown as solid blue and dashed red lines, 
respectively.}  
\label{fig:wfn}
\end{center} 
\end{figure}

The first example is a Hamiltonian with a moderately sized barrier;
its parameters are  $V_0=1$ in energy units of
$E_q$ and $\sigma = 2$ in  length units of $s$.  This barrier height 
is well within the domain of acceptable energies.  Also 
the barrier curvature parameter
expressed as a harmonic oscillator energy $\hbar \omega$ is within
the domain. 
Fig. \ref{fig:wfn} displays the calculated wave function for an
incident energy just at the barrier top, $E = V_0$.  The  numerical
parameters are
$\Delta q   = (2/3) 5^{1/2} \,s $,
$N_{od} = 2$, and $N_{DB} = 42$.   The dimension of the matrices
$N_{DB}$ is much larger than necessary; the purpose is to exhibit the
plane-wave character of the solution outside the barrier region.
The points show the real and imaginary parts 
of the scattering wave function in
the $q$-representation calculated as
\be
\Psi(q) = \sum_if_i\psi_i(q)=\sum_i f_i \exp(-(q-q_i)^2/2s^2).
\ee
The wave function on the right-hand side is clearly a
traveling wave of the form $e^{i k x}$ with $k>0$ as required for
an outgoing flux. The wave on the other side has both incoming and
outgoing components that almost add together for a standing wave pattern.
This is somewhat deceptive.  A pure standing wave would have equal
amplitudes of incoming and outgoing components implying a reflection
probability of one.  The actual reflection  probability in this
example is close to $1/2$, the expected value in 
the Hill-Wheeler formula.

\begin{figure}[htb] 
\begin{center} 
\includegraphics[trim= 0   0  0 0,clip=true,width=\columnwidth]{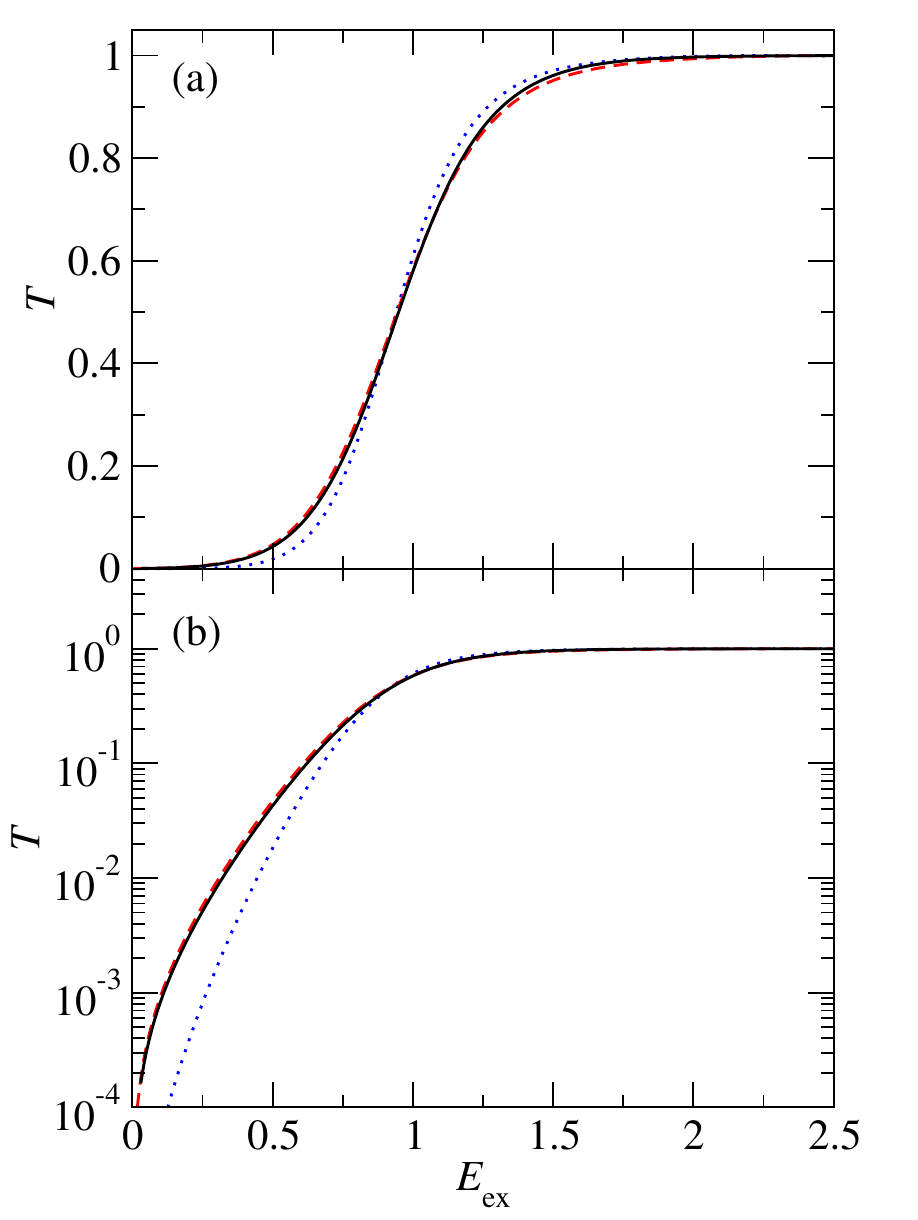}
\caption{Transmission probabilities $T$ for barrier crossing as a function
of excitation energy $E_{\rm ex}$,  plotted in linear (upper panel) and  
logarithmic 
(lower panel) scales.  The barrier has the form Eq. (\ref{eq:V}) with
$V_0 = E_q $ and $\sigma = 2\, s$.  Shown are the transmission probabilities
from the  Schr\"odinger equation  (solid black line), the 
discrete-basis equation in the tridiagonal approximation  (blue dotted 
line), and the discrete-basis equation with next-to-nearest neighbor
interactions (red dashed line). 
\label{fig:TvsE}} 
\end{center} 
\end{figure}

We next compare the energy dependence of 
$T_{DB}$ with Schr\"odinger solutions, taking
the same barrier parameters as before. The 
numerical parameters  are set to $N_{DB} = 10$  and $\Delta q = 5^{1/2} s$
to show what can be achieved in a small space.
The Hamiltonian is defined on a range of $q$ that is long
enough to cover the barrier region and meet the
criteria for plane-wave behavior near the end points 
The calculated $T_{DB}(E)$ is shown in Fig. \ref{fig:TvsE}, 
both in linear and logarithmic scales,  together with that obtained by
solving the Schr\"odinger equation\footnote{See the Supplementary
Material for computer codes to perform
the calculations.}. 
The figure shows that the discrete-basis approach with $N_{od}=2$ is
in excellent agreement with the Schr\"odinger equation, even at deep subbarrier 
energies.  Also,
the more economical $N_{od} =1$ treatment with a somewhat larger
mesh spacing is useful, given that the microscopic 
nuclear Hamiltonians in current use have limited predictive power. 

\begin{figure}[htb] 
\begin{center} 
\includegraphics[clip=true,width=\columnwidth]{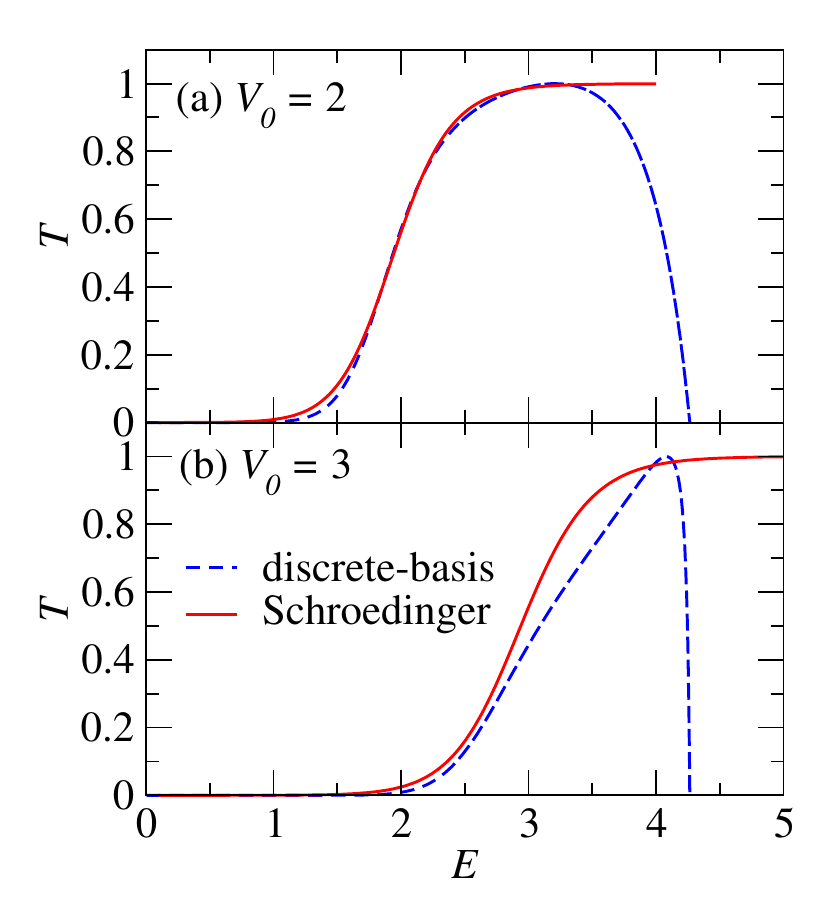} 
\caption{Transmission coefficients for $V_0 = 2 E_q$ (upper panel) and
$V_0 = 3 E_q$ (lower panel).  Blue dashed:  Schr\"odinger equation;
red solid: discrete-basis method with $N_{od} = 1$ and $\Delta_q = 5^{1/2}$.  
\label{fig:T2T3}} 
\end{center} 
\end{figure}

\begin{figure}[htb] 
\begin{center} 
\includegraphics[trim= 0   0  0  0,clip=true,width=\columnwidth]{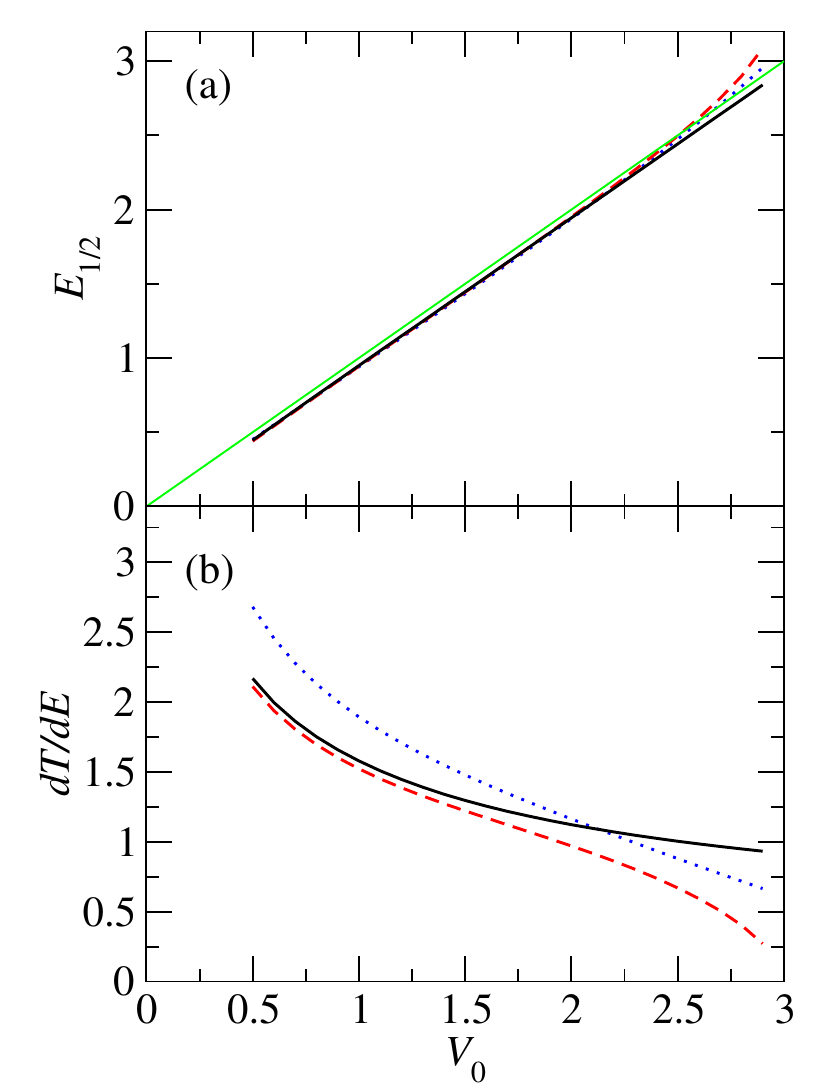}
\caption{Panel a): $E_{1/2}$ vs $V_0$ for the Schr\"odinger Hamiltonian
(solid black) compared to the discrete-basis Hamiltonian (dashed blue). 
$E_{1/2} = V_0$ on the dotted black line.  Panel b):  $dT/dE$ at $E_{1/2}$
vs $V_0$.  Schr\"odinger results are shown by the solid black line.  The
discrete-results are shown by dotted blue and the dashed red lines for
$N_{od}=1$ and 2, respectively.  Other parameters
are the same as in Fig. \ref{fig:TvsE}.
\label{fig:EhalfdT}} 
\end{center} 
\end{figure}

We now examine limits of the discrete-basis approach for higher barriers.
Fig. \ref{fig:T2T3}  shows $T_{DB}(E)$ for $V_0 = 2 $ and $3 E_q$.
At both barrier heights  the discrete-basis Hamiltonian is not useful
above $E \sim 3 E_q$.  
For a more quantitative assessment of the performance we examine
energies $E_{1/2}$ at which the transmission probability reaches
$1/2$, i.e. $ T_{DB}(E_{1/2}) = 1/2$,  and its slope $ dT /dE$ at that energy.
These are presented in Fig. \ref{fig:EhalfdT}.  The parameters are the
same as before except for $N_{od}$.
One sees that $E_{1/2}$ is quite accurate up to $V_0 = 3 E_q$.  But this
is somewhat misleading because the full $T_{DB}(E)$ curve does not reach close
to $T=1$ at higher energies.  From the lower panel one sees that the
transmission probability rises somewhat more sharply for the tridiagonal 
Hamiltonian than for the Schr\"odinger $H_s$ in most of safe region of energies. 
However, the $N_{od}=2$ treatment is quite accurate at low energies.

\section{Conclusion}
  At a purely
phenomenological level, the one-dimensional Hamiltonian proposed by Hill and
Wheeler leads to a simple formula  that remains an integral part of 
fission phenomenology\cite{ko23,ca09}.   But fission theory at a
microscopic level relies on a many-particle formalism to create a
matrix Hamiltonian or to determine the
parameters of a Schr\"odinger  Hamiltonian.
This work has shown
that the usual procedure for building a CI basis can mimic the
Schr\"odinger approach quite well.  However, there is a important restriction on
its applicability.
Namely, the barrier height cannot be much higher than a 
few times the zero-point energies of the configurations as 
given by $E_q$ in Eq. (\ref{eq:Eq}). In Ref. \cite{BH2023} the
functional form of the equation is verified for a few
barrier-top configurations finding $E_q$ in the range\footnote{$E_q= h_2/2$ in the notation of
Ref. \cite{BH2023}.}$ 1.5- 2 $ MeV. Whether this is too small depends on
the details of how the paths to a transition-state start
out.  
The barrier
heights of well-known fissile nuclei are of the order of $6$ MeV above the
ground-state
energy, somewhat outside the reach of
the present  approach.  However, CI configurations at more
favorable energies in the compound nucleus might 
be diabatically connected to the
transition states, giving more scope to the method.

Two ways come to mind for  increasing the space of higher energy
excitations in the collective variables.  In reaction theory of small
clusters, excitations of their  center-of-masses can be introduced by
algebraic operators in the harmonic oscillator representation\cite{kr19}.  
However, that representation may not be practical for heavy nuclei.  Another
approach would be to include momentum constraints at the mean field level
to increase the energies with respect to collective coordinates
\cite{re87,hi22}.  This may
have been explored for small-amplitude shape changes, but to our knowledge
has not been implemented in codes for generating a CI basis  in heavy nuclei.

If the space were large enough to use the discrete-basis approach
with confidence, a fundamental question in fission theory could
be addressed.  Namely, what  is the relative 
importance of collective flow versus diffusive flow in large-amplitude shape
changes?  
In one extreme, the shape changes go mainly through collective 
coordinates that lead to  a Schr\"odinger equation in one- or a
few-dimensions.  In another extreme, the shape
changes come about as a random walk through non-collective intermediate
configurations.
There are compelling arguments that diffusive flow dominates at
energies much higher than the barrier \cite{bu92}. On the other hand theory 
based on an adiabatic collective coordinate does quite well at the
far subbarrier energies associated with spontaneous fission 
\cite{schunck2016,whitepaper}.
It seems to us that some form of a CI
approach is needed for treating both mechanisms on the same footing.

As a final remark, the present  analysis is  based on the factorization 
hypothesis contained in Eq. (\ref{eq:psiq}) and leading to the dynamics
derived from Eq. (\ref{eq:Eq}).   The conclusions regarding the adequacy of the 
generated space of configurations would certainly be affected.

\begin{acknowledgments}
We thank T. Kawano and L.M. Robledo for helpful comments on the manuscript.
This work was supported in part by
JSPS KAKENHI Grant Numbers JP19K03861 and JP23K03414.
\end{acknowledgments}

\section{Appendix}

We derive here the equation for the wave function satisfying
plane-wave boundary conditions of the discrete-basis Hamiltonian.
The outgoing wave function has the form
$ f_m = C r^m$  where  $r = \exp( i k \Delta x)$ and  $C$ is an
arbitrary constant.  The equation
for row $m_0$  in the matrix $H'$  reads
\be
 r^{m_0}\sum_{m = -N_{od}}^{N_{od}} h'_{|m|}  C r^m  = 0
\ee
providing that the Hamiltonian matrix elements are beyond the
range of the potential $V$ and that the row is within
$ N_{od}  < m_0 \leq N_{\rm DB} -N_{od} $.  The
missing terms  of rows $N_{DB}-N_{od}< m_0\leq N_{DB}$ are 
added to the diagonal element
in that row,
\be
\Delta H'_{m_0,m_0}  =  \sum_{m=m'}^{N_{od}}  h'_m r^m 
\ee   
where $m' = N_{DB} -m_0 + 1 $.
To deal with the missing entries in the beginning rows, we  
consider the incoming channel amplitude $f_0$ on the site adjacent
to the first site $q_1$ in the matrix Hamiltonian and possibly others
if $N_{od} >1$.  The contribution of $f_0$
 is missing from rows of $H'\psi$ in the range
$ m_0 \le N_{od}$.
Only the term $h'_1 f_0$ is missing in the last of these rows.
It is separated out as an
inhomogenous term in the Hamiltonian equation.  

For $N_{od} =1 $  the matrix $H'$ is tridiagonal and 
the equation to be solved is
$(\mat{H}' + \mat{\Delta H}')\vec{f} = - h_1 \vec{v}$  
with vector $\vec{v} =  (f_0,0,\cdots, 0)$.  For the numerical
solution, one can set  $f_0= 1$ and determine the
rest of the wave function by matrix inversion
as in Ref. \cite{BH2023}.

The wave function around the first site  will have outgoing
as well as incoming components for the full Hamiltonian
with a barrier $V$.
The amplitudes of incoming and reflected components 
$(a_{\rm in},a_{\rm out})$ can be extracted from the wave function
amplitudes at $f_0$ and $f_1$,

\be
\label{H} 
\left(\begin{matrix} 
a_{\rm in} \cr
a_{\rm out}  \cr
\end{matrix} 
\right)
= \frac{1}{r -r^{-1}}
\left[\begin{matrix} r   & -1  \cr
 -r^{-1} & 1  \cr
\end{matrix}\right] 
\left(\begin{matrix} f_1 \cr
f_0 \cr
\end{matrix}\right).
\ee
This is the end of the story for $N_{od} =1 $.

For $N_{od} >1 $, there are other incomplete rows in the
Hamiltonian matrix requiring amplitudes
$f_0,..,f_{-N_{od}+1}$.  These can be
determined from $(a_{\rm in},a_{\rm out})$ as
\begin{eqnarray}
f_m &=&  a_{\rm in} r^{-(1-m)} + a_{\rm out} r^{(1-m)},   \\
& = & \frac{1}{r-r^{-1}}\left( (r^{-m} -r^m)f_1 
+(r^{-m+1}- r^{m-1}) f_0\right) .  \nonumber  \\
\end{eqnarray} 
The coefficients of terms 
with $f_i$ on the second line
are added to the Hamiltonian matrix element $\mat{H}'_{1,m_0}$ while
the terms with $f_0$ are added to vector $\vec{v}$ in Eq. (\ref{eq:H''v}).  
In detail, the matrix elements added to $\mat{H}'$ for $N_{od} =2$ are
\begin{eqnarray}
&&\mat{\Delta H}_{11} = - h'_2 \left(1/(1- r^2) + 1/(1-r^{-2})\right),  \\
&&\mat{\Delta H}_{N_{DB},N_{DB}} = h'_1 r + h'_2 r^2, \\
&&\mat{\Delta H}_{N_{DB}-1,N_{DB}-1}= h'_2 r^2,
\end{eqnarray}
and the nonzero components of the vector $v$ are  
\begin{eqnarray}
v_1 &=& h'_1 + h'_2\left(r^{-1}/(1-r^2) + r/(1-r^{-2})\right), \\
v_2 &=& h'_2.
\end{eqnarray}

The final inhomogenious equation to be solved is
\be
(\mat{H}' + \mat{\Delta H}') \vec{f}  = - \vec{v} f_0
\ee
with an arbitrary nonzero $f_0$.

\end{document}